\documentclass[prc,twocolumn]{revtex4}

\usepackage{amssymb}

\newcommand{\be}{\begin{equation}}
\newcommand{\ee}{\end{equation}}
\newcommand{\ba}{\begin{eqnarray}}
\newcommand{\ea}{\end{eqnarray}}
\newcommand{\ban}{\begin{eqnarray*}}
\newcommand{\ean}{\end{eqnarray*}}

\begin{document}

\title{Quasi-linear transport approach to equilibration 
of quark-gluon plasmas}

\author{Stanis\l aw Mr\' owczy\' nski}

\affiliation{Institute of Physics, Jan Kochanowski University, 
25-406 Kielce, ul. \'Swi\c etokrzyska 15, Poland}
\affiliation{So\l tan Institute for Nuclear Studies, 00-681 Warsaw, 
ul. Ho\. za 69, Poland}

\author{Berndt M\"uller}

\affiliation{Department of Physics \& CTMS, Duke University, 
Durham, NC 27708, USA}

\date{28-th January 2010}

\begin{abstract}

We derive the transport equations of quark-gluon plasma in the 
quasi-linear approximation. The equations are either of the 
Balescu-Lenard or Fokker-Planck form. The plasma's dynamics is 
assumed to be governed by longitudinal chromoelectric fields. 
The isotropic plasma, which is stable, and the two-stream system, 
which is unstable, are considered in detail. A process of 
equilibration is briefly discussed in both cases. The peaks 
of the two-stream distribution are shown to rapidly dissolve 
in time.

\end{abstract}

\maketitle


\section{Introduction}


The quark-gluon plasma (QGP), which is produced in relativistic
heavy-ion collisions, is believed to be equilibrated within a time interval
of order of 1 fm/$c$ or even shorter \cite{Heinz:2004pj}. Such 
a fast equilibration is naturally explained assuming that the quark-gluon 
plasma is strongly coupled \cite{Bannur:1998nq,Gyulassy:2004zy,Shuryak:2008eq}. 
Then, scattering processes are very frequent and relaxation times are short. 
However, the theory of high-energy density QCD \cite{McLerran:1993ni}
suggests that due to the existence of a large momentum scale $Q_s$,
at which the gluon density saturates, the plasma is rather weakly 
coupled at the early stage of the collision because of asymptotic 
freedom.  Experimental data on jet quenching indicate that the coupling 
constant $\alpha_s\leq 0.3$ \cite{Qin:2007rn,Nagle:2008fw}, even though 
the value assumes averaging over the whole evolution of the QCD medium 
created in the relativistic heavy-ion collision. Thus, the question 
arises how fast the weakly interacting plasma equilibrates. Due to 
anisotropic momentum distributions the early stage plasma is 
unstable with respect to the chromomagnetic plasma modes. The 
instabilities isotropize the system and thus speed up the process 
of its equilibration. The scenario of the instabilities-driven 
isotropization is reviewed in \cite{Mrowczynski:2005ki}. However, the 
complete evolution of the plasma momentum distribution is now 
accessible only by numerical simulations 
\cite{Dumitru:2005gp,Dumitru:2006pz,Arnold:2005ef}.

The transport theory of a weakly coupled quark-gluon plasma has been 
studied since 1980s when the kinetic equations in the mean-field 
approximation were derived \cite{Elze:1989un,Mrowczynski:1989np}.
Although the mean-field dynamics is rather simplified, the 
equations are still difficult to solve due to their non-linear 
structure. If one is interested in small deviations from 
equilibrium or any other homogeneous and stationary state, the 
equations can be linearized and then solved. The mean-field
transport theory, which is linearized in small deviations from 
equilibrium, is now well understood, for a review see 
\cite{Blaizot:2001nr}. It is known to be equivalent to the 
effective QCD in the hard-thermal loop approximation. The 
linearized transport theory around any homogeneous and 
stationary but non-equilibrium plasma state was also worked out 
and the connection with the diagrammatic hard loop approximation 
was  established \cite{Mrowczynski:2000ed,Mrowczynski:2004kv}. 
Numerous problems of the theory of the quark-gluon plasma were 
successfully resolved within the hard loop approach. For example, 
a systematic method to eliminate infrared divergences, which 
plague perturbative calculations, was developed, 
see the reviews \cite{Thoma:1995ju,Kraemmer:2003gd}. 

However, various questions cannot be addressed within the 
mean-field theory. For example, transport coefficients are 
then formally infinite. Thus, there were numerous efforts 
to derive transport equations of quark-gluon plasma which hold 
beyond the hard loop approximation
\cite{Selikhov:1991br,Selikhov:1993ns,Selikhov:1994xn,Zhang:1995si,Zheng:1997mh, Bodeker:1998hm,Bodeker:1999ey,Arnold:1998cy,Arnold:1999jf,Arnold:1999uy,Litim:1999ns,Litim:1999id,ValleBasagoiti:1999pg,Blaizot:1999xk,Markov:2000pb,Markov:2002su,Markov:2003rk,Markov:2005qe,Markov:2005gn,Markov:2006wp,Akkelin:2008bq}.
These efforts were mostly concerned with the transport properties
of an {\em equilibrium} quark-gluon plasma. Our motivation is rather 
different. We are interested in {\em equilibration} of quark-gluon 
plasma, in particular in the equilibration of the system which is 
initially unstable. Thus, we intend to study how fluctuating deviations 
from a quasi-stationary non-equilibrium state influence the system's 
bulk or average momentum distribution. This effect of back reaction is 
particularly important in the case of unstable systems. The linear 
response theory describes how unstable modes initially grow in the 
presence of a non-equilibrium momentum distribution, but it says 
nothing on how the modes modify the plasma momentum distribution. Thus, 
the problem of equilibration cannot be addressed in such a theory. 

Our objective here is to derive the transport equations where the 
bulk distribution function slowly evolves due to the interaction 
with fluctuating chromodynamic fields. We actually consider only 
a simplified problem of QGP in a self-consistently generated 
{\em longitudinal} chromoelectric field. This simplification is
not much needed for isotropic plasma but it appears crucial to 
study anisotropic systems. Taking into account only the longitudinal 
chromoelectric field, we obtain the transport equations of the 
Fokker-Planck or Balescu-Lenard form which describe the 
effect of back reaction. A similar, but incomplete effort was undertaken 
by Akkelin \cite{Akkelin:2008bq}. The derivation presented here 
closely follows the procedure developed for the 
electromagnetic plasma, where it is known as the quasi-linear theory 
or the theory of a weakly turbulent plasma \cite{Ved61,Ved63,LP81}. 
The theory assumes that the distribution function of plasma 
particles can be decomposed into a large but slowly varying regular 
part and a small fluctuating or turbulent one which oscillates fast. 
The average over the statistical ensemble of the turbulent part is 
assumed to vanish and thus the average of the distribution function 
equals its regular part. The turbulent contribution to the distribution 
function obeys the collisionless transport equation while the transport 
equation of the regular part, which is of our main interest here, is 
determined by the fluctuation spectra. The fluctuations of 
chromodynamic fields, which are used to derive the quasi-linear 
transport equations, were studied in \cite{Mrowczynski:2008ae}
where stable and unstable plasma states were considered. 

The Fokker-Planck equation derived here is somewhat
similar to the equation obtained in 
\cite{Asakawa:2006tc,Asakawa:2006jn}. It was used there to show 
that the chromomagnetized quark-gluon plasma exhibits an anomalous 
shear viscosity, as presence of the domains of chromomagnetic 
field leads to the momentum transport in the plasma.

Our paper is organized as follows. In Sec.~\ref{sec-preliminaries}
we present the QGP transport equations; the notation and conventions 
are introduced. The decomposition of the distribution functions into 
the regular and turbulent parts is discussed in 
Sec.~\ref{sec-reg-fluc}. The explicit expressions of the fluctuating 
distribution functions which obey collisionless transport equations 
are derived in Sec.~\ref{sec-sol-lin}. A general form of the equations 
of the regular distribution functions is found here as well. Further
discussion splits into two parallel parts: Sec.~\ref{sec-iso} is 
devoted to the stable isotropic plasma while in 
Sec.~\ref{sec-2-streams} the unstable two-stream system is discussed. 
Although we neglect transverse chromodynamic fields, the collision 
terms of transport equations, which are found here for an isotropic 
plasma, are very similar to those derived in 
\cite{Selikhov:1991br,Selikhov:1994xn,Bodeker:1999ey,Bodeker:1999ey,Arnold:1998cy,Litim:1999ns,Litim:1999id,ValleBasagoiti:1999pg,Blaizot:1999xk,Markov:2000pb}.
As an application of the transport equations we derived, a process 
of equilibration of the isotropic plasma and of the two-stream 
system is discussed. The paper closes with a summary of our 
considerations and outlook. 


\section{Preliminaries}
\label{sec-preliminaries}


The transport theory of a quark-gluon plasma, which forms the basis 
of our analysis, is formulated in terms of particles and classical 
fields. The particles - quarks, antiquarks and gluons - should
be understood as sufficiently hard quasiparticle excitations of
quantum fields of QCD while the classical fields are highly populated 
soft gluonic modes. An excitation is called ``hard'', when its 
momentum in the equilibrium rest frame is of order of the temperature 
$T$, and it is called ``soft'' when the momentum is of order $gT$ with 
$g$ being the coupling constant. Since we consider a weakly coupled 
quark-gluon plasma, the coupling constant is assumed to be small 
$g \ll 1$. In our further considerations the quasiparticles are 
treated as classical particles obeying Boltzmann statistics but 
the effect of quantum statistics can be easily taken into account.

The transport equations of quarks, antiquarks and gluons are 
assumed to be of the form
\ba
\nonumber
{\cal D}\, Q(t,{\bf r},{\bf p})
- {1 \over 2}
\{{\bf F}(t,{\bf r}) , \nabla_p Q(t,{\bf r},{\bf p}) \}
&=& 0 \;, 
\\ 
\label{transport-eq}
{\cal D}\, \bar Q(t,{\bf r},{\bf p})
+ {1 \over 2}
\{{\bf F}(t,{\bf r}) , \nabla_p \bar Q(t,{\bf r},{\bf p}) \}
&=& 0 \;, 
\\ \nonumber
{\cal D}\, G(t,{\bf r},{\bf p})
- {1 \over 2}
\{{\bf F}(t,{\bf r}) , \nabla_p G(t,{\bf r},{\bf p}) \}
&=&  0\;.
\ea
The (anti-)quark distribution functions $Q(t,{\bf r},{\bf p})$ 
and $\bar Q(t,{\bf r},{\bf p})$, which are $N_c\times N_c$ hermitean 
matrices, belong to the fundamental representation of the SU($N_c$) 
group, while the gluon distribution function $G(t,{\bf r},{\bf p})$, 
which is a $(N_c^2 -1) \times (N_c^2 -1)$ matrix, belongs to the 
adjoint representation. The distribution functions depend on the 
time ($t$), position (${\bf r}$) and momentum (${\bf p}$) variables. 
There is no explicit dependence on the time-like ($\mu=0$) component 
of the four-vector $p^\mu$ as the distribution functions are assumed 
to be non-zero only for momenta obeying the mass-shell constraint 
$p^\mu p_\mu =0$. Because the partons are assumed to be massless, 
the velocity ${\bf v}$ equals ${\bf p}/E_{\bf p}$ with 
$E_{\bf p}=|{\bf p}|$. ${\cal D} \equiv D^0 +{\bf v} \cdot {\bf D}$ 
is the covariant substantive derivative given by the covariant 
derivative which in the four-vector notation reads 
$D^\mu \equiv \partial^\mu - ig [A^\mu(x),\cdots \;]$ with 
$A^\mu(x)$ being the chromodynamic potential. The mean-filed terms
of the transport equations (\ref{transport-eq}) are expressed 
through the color Lorentz force ${\bf F}(t,{\bf r}) \equiv 
g\big({\bf E}(t,{\bf r}) + {\bf v} \times {\bf B}(t,{\bf r})\big)$.
The chromoelectric ${\bf E}(t,{\bf r})$ and chromomagnetic
${\bf B}(t,{\bf r})$ fields belong to either the fundamental or 
adjoint representation. To simplify the notation we use the same 
symbols ${\cal D}$, $D^0$, ${\bf D}$, ${\bf E}$, and ${\bf B}$ for 
a given quantity in the fundamental or adjoint representation. 
The symbol $\{\dots , \dots \}$ denotes the anticommutator.

The collision terms are neglected in the transport equations 
(\ref{transport-eq}). The collisionless equations are applicable 
in three physically different situations: when the distribution 
function is of (local) equilibrium form; when the timescale of 
processes of interest is much shorter than the average temporal 
separation of parton collisions; and when the system dynamics is 
dominated by the mean field. In our study we refer to all three 
situations. When the equilibration of isotropic plasma is discussed, 
it is crucial that the collision terms vanish in local equilibrium. 
In the case of unstable two-stream system, the effects of collisions 
can be initially neglected, as the growth of unstable modes is very 
fast. Later on, the strong fields become mostly responsible for the
system's evolution 

The transport equations are supplemented by the Yang-Mills equations
describing a self-consistent generation of the chromoelectric and 
chromomagnetic fields. The equations read
\ba
\nonumber
{\bf D} \cdot {\bf E}(t, {\bf r}) &=&  \rho (t, {\bf r}) 
\\  \nonumber
{\bf D} \cdot {\bf B}(t, {\bf r}) &=& 0 \;, 
\\ 
\label{YM-eqs-x}
{\bf D} \times {\bf E}(t, {\bf r}) &=& - 
D_0 {\bf B}(t, {\bf r}) \;,
\\  \nonumber
{\bf D} \times {\bf B}(t, {\bf r}) &=& 
{\bf j}(t, {\bf r}) + D_0 {\bf E}(t, {\bf r}) \;,
\ea
where the color four-current $j^\mu =(\rho , {\bf j})$ in the adjoint
representation equals
\ba
\label{current-def}
\nonumber
j^\mu_a (t,{\bf r}) &=& - g \int {d^3 p \over (2\pi)^3} \,
\frac{p^\mu}{E_{\bf p}}
{\rm Tr}\Big[ T^a G(t,{\bf r},{\bf p}) 
\\ && \qquad
+ \tau^a \big(Q (t,{\bf r},{\bf p}) - \bar Q(t,{\bf r},{\bf p}) \big)\Big] \;,
\ea
where $\tau^a$, $T^a$ with $a = 1, ... \, ,N_c^2-1$ are the SU($N_c$)
group generators in the fundamental and adjoint representations,
normalized as ${\rm Tr}[\tau^a \tau^b] = \frac12 \delta^{ab}$ and
${\rm Tr}[T^a T^b] = N_c \delta^{ab}$. The set of transport equations
(\ref{transport-eq}) and Yang-Mills equations (\ref{YM-eqs-x}) 
is covariant with respect to ${\rm SU}(N_c)$ gauge transformations. 


\section{Regular and Fluctuating Quantities}
\label{sec-reg-fluc}


We assume that the chromodynamic fields and distribution functions
which enter the set of transport equations can be decomposed into 
the regular and fluctuating components. The quark distribution 
function is thus written down as
\be
\label{reg-fluc}
Q(t,{\bf r},{\bf p}) = \langle Q(t,{\bf r},{\bf p}) \rangle 
+ \delta Q(t,{\bf r},{\bf p})
\;,
\ee
where $\langle \cdots \rangle$ denotes ensemble average; 
$\langle Q(t,{\bf r},{\bf p}) \rangle$ is called the regular part 
while $\delta Q(t,{\bf r},{\bf p})$ is called the fluctuating or 
turbulent one. It directly follows from Eq.~(\ref{reg-fluc}) that 
$\langle \delta Q \rangle =0$. The regular contribution is assumed 
to be white, and it is expressed as   
\be
\label{whiteness}
\langle Q(t,{\bf r},{\bf p}) \rangle  = 
n (t,{\bf r},{\bf p}) \, I \;, 
\ee
where $I$ is the unit matrix in color space. Since the 
distribution function transforms under gauge transformations as 
$Q \rightarrow U\, Q U^{-1}$, where $U$ is the transformation 
matrix, the regular contribution of the form (\ref{whiteness})
is gauge independent. We also assume that
\be
\label{quasi-linear-conditions-1}
|\langle Q \rangle | \gg |\delta Q| \;, \qquad
|\nabla_p \langle Q \rangle | \gg | \nabla_p \delta Q| \;,
\ee
but at the same time 
\be
\label{quasi-linear-conditions-2}
\left|\frac{\partial \delta Q }{\partial t} \right| 
\gg \left|\frac{\partial \langle Q \rangle}{\partial t} \right| 
\;, \qquad
| \nabla \delta Q| \gg |\nabla \langle Q \rangle | \;.
\ee
Analogous conditions are assumed for the antiquark and gluon distribution
functions. What concerns the chromodynamic fields, we assume in accordance
with (\ref{whiteness}) that their regular parts vanish and thus 
\be
\langle {\bf E}(t,{\bf r}) \rangle = \langle {\bf B}(t,{\bf r}) \rangle = 0 \;.
\ee

We substitute the distribution functions (\ref{reg-fluc}) 
into the transport equations and the Yang-Mills equations and linearize
the equations in the fluctuating contributions. The linearized transport 
and Yang-Mills equations remain rather complex. Therefore, we discuss 
here a simplified problem: we consider a QGP in the presence of turbulent 
{\em longitudinal} chromoelectric fields, but neglect the chromomagnetic
and transverse chromoelectric fields. This simplification can be avoided 
for an isotropic plasma but it is needed, as explained below, to make 
progress on an analytical treatment for anisotropic systems which are 
our main interest here. The simplified transport equations then read
\ba
\nonumber 
{\cal D}\, \delta Q(t,{\bf r},{\bf p})
- g {\bf E}(t,{\bf r}) \cdot \nabla_p n(t,{\bf r},{\bf p})
&=& 0 \;, 
\\   
\label{trans-eq-lin}
{\cal D}\, \delta \bar Q(t,{\bf r},{\bf p})
+ g {\bf E}(t,{\bf r}) \cdot \nabla_p \bar n(t,{\bf r},{\bf p})
&=& 0 \;, 
\\ \nonumber
{\cal D}\, \delta G(t,{\bf r},{\bf p})
- g {\bf E} (t,{\bf r})\cdot \nabla_p n_g(t,{\bf r},{\bf p})
&=&  0\;,
\ea
where 
${\cal D} \equiv  \frac{\partial}{\partial t} + {\bf v}\cdot \nabla $
denotes from now on the material (not covariant) derivative.

The equation describing the self-consistent generation of a 
longitudinal chromoelectric field is
\be
\label{YM-eqs-x-lin}
\nabla \cdot {\bf E}_a(t, {\bf r}) = \rho_a (t, {\bf r})=
- g \int {d^3 p \over (2\pi)^3} \, \delta N_a(t,{\bf r},{\bf p})
\;,
\ee
where 
\ba
\label{N-def}
\nonumber
\delta N_a(t,{\bf r},{\bf p}) &\equiv&
{\rm Tr}\big[\tau^a \big(\delta Q (t,{\bf r},{\bf p}) 
- \delta \bar Q(t,{\bf r},{\bf p}) \big)
\\ && \hspace{2cm}
+ T^a \delta G(t,{\bf r},{\bf p}) \big]\;.
\ea
The linearized equations are formally Abelian but they include
a fundamentally non-Abelian effect, {\em i.~e.} the gluon contribution 
to the color current. Therefore, the gluon-gluon coupling is
partly taken into account. The linearized Yang-Mills equation
corresponds to the multi-component electrodynamics of $N_c$ charges
(in the so-called Heaviside-Lorentz system of units). The equations, 
however, are no longer manifestly covariant with respect to 
${\rm SU}(N_c)$ gauge transformations. Nevertheless, our final results 
are gauge independent. 

We now substitute the distribution functions (\ref{reg-fluc}) into the 
transport equations (\ref{transport-eq}). 
Instead of linearizing the equations in the fluctuating contributions, 
we take the ensemble average of the resulting equations and trace 
over the color indices. Thus we get
\ba
\nonumber 
{\cal D}\, n - \frac{g}{N_c}{\rm Tr} 
\langle {\bf E} \cdot \nabla_p \delta Q \rangle 
&=& 0 \;, 
\\ 
\label{trans-eq-reg}
{\cal D}\, \bar n + \frac{g}{N_c}{\rm Tr} 
 \langle {\bf E} \cdot \nabla_p \delta \bar Q \rangle 
&=& 0 \;, 
\\ \nonumber 
{\cal D}\, n_g - \frac{g}{N_c^2 - 1}{\rm Tr} 
 \langle {\bf E} \cdot \nabla_p \delta G \rangle 
&=&  0 \;.
\ea
Since the regular part of distribution function is assumed to be 
color neutral, see Eq.~(\ref{whiteness}), the terms of the form
${\rm Tr}[\langle {\bf E} \cdot \nabla_p n \rangle]$ vanish 
because the field ${\bf E}$ is traceless. The trace over color 
indices also cancels the terms originating from covariant 
derivatives like ${\rm Tr}\langle [A^\mu,\delta Q] \rangle$. 
We finally note that the regular distribution function $n$ 
is gauge independent and so is 
${\rm Tr} \langle {\bf E} \cdot \nabla_p \delta Q \rangle$.


\section{Solution of the linearized equations}
\label{sec-sol-lin}


Due to the condition (\ref{quasi-linear-conditions-2}), the space-time
dependence of the regular distribution functions is neglected 
in the linearized transport equations (\ref{trans-eq-lin}) and the 
equations become easily solvable. We solve Eq.~(\ref{trans-eq-lin}) 
with the initial conditions
\ba
\delta Q(t=0,{\bf r},{\bf p}) &=& \delta Q_0({\bf r},{\bf p}) \; ,
\nonumber 
\\
\delta \bar Q(t=0,{\bf r},{\bf p}) &=& \delta \bar Q_0({\bf r},{\bf p}) \; ,
\\ \nonumber 
\delta G(t=0,{\bf r},{\bf p}) &=& \delta G_0({\bf r},{\bf p}) \;,
\ea
using the one-sided Fourier transformation defined as
\be
f(\omega,{\bf k}) = \int_0^\infty dt \int d^3r 
e^{i(\omega t - {\bf k}\cdot {\bf r})} f(t,{\bf r}) \;.
\ee
The inverse transformation is 
\be
f(t,{\bf r}) = \int_{-\infty +i\sigma}^{\infty +i\sigma}
{d\omega \over 2\pi} \int {d^3k \over (2\pi)^3} 
e^{-i(\omega t - {\bf k}\cdot {\bf r})} f(\omega,{\bf k}) \;,
\ee
where the real parameter $\sigma > 0$ is chosen in such a
way that the integral over $\omega$ is taken along a straight
line in the complex $\omega-$plane, parallel to the real axis, 
above all singularities of $f(\omega,{\bf k})$. We note that 
\ba
 -i\omega f(\omega,{\bf k}) &=& f(t=0,{\bf k}) 
\\  \nonumber 
&+& \int_0^\infty dt \int d^3r 
e^{i(\omega t - {\bf k}\cdot {\bf r})}
{\partial f(t,{\bf r}) \over \partial t} \;.
\ea

The linearized transport equations (\ref{trans-eq-lin}), which are 
transformed by means of the one-sided Fourier transformation, are 
solved as
\ba
\nonumber
\delta Q(\omega,{\bf k},{\bf p})
&=& i\frac{ g {\bf E}\cdot
\nabla_p n({\bf p}) +  \delta Q_0({\bf k},{\bf p})}
{\omega - {\bf k}\cdot {\bf v}}
 \;,
\\ 
\label{solution1}
\delta \bar Q(\omega,{\bf k},{\bf p})
&=& i \frac{ g {\bf E}\cdot
\nabla_p \bar n ({\bf p}) -  \delta \bar Q_0({\bf k},{\bf p})}
{\omega - {\bf k}\cdot {\bf v}} \;,
\\ \nonumber
\delta G(\omega,{\bf k},{\bf p})
&=& i \frac{ g {\bf E}\cdot
\nabla_p n_g({\bf p}) +  \delta G_0({\bf k},{\bf p})}
{\omega - {\bf k}\cdot {\bf v}}
 \;.
\ea
We note that the color electric field ${\bf E}(\omega,{\bf k})$ retains 
its full frequency and wave number dependence in these equations.
Inverting the one-sided Fourier transformation, one finds
the solutions of linearized transport equations as
\begin{widetext}
\ba
\nonumber
\delta Q(t,{\bf r},{\bf p}) &=&
g \int_0^t dt' \:
{\bf E}\big(t', {\bf r}-{\bf v} (t-t')\big) 
\cdot \nabla_p n({\bf p})
+ \delta Q_0({\bf r}-{\bf v} t,{\bf p}) \;,
\\ 
\label{sol-x}
\delta \bar Q(t,{\bf r},{\bf p}) &=&
-g \int_0^t dt' \:
{\bf E}\big(t', {\bf r}-{\bf v} (t-t')\big) 
\cdot \nabla_p \bar n({\bf p})
+ \delta \bar Q_0({\bf r}-{\bf v} t,{\bf p}) \;,
\\ \nonumber 
\delta G(t,{\bf r},{\bf p}) &=&
g \int_0^t dt' \:
{\bf E}\big(t', {\bf r}-{\bf v} (t-t')\big) 
\cdot \nabla_p n_g({\bf p})
+ \delta G_0({\bf r}-{\bf v} t,{\bf p}) \;,
\ea
where we assumed that ${\bf E}(\omega,{\bf k})$ is an analytic function
of $\omega$. With the help of solutions (\ref{sol-x}), the force terms 
in the transport equations (\ref{trans-eq-reg}) become
\ba
\nonumber
\langle {\bf E}(t, {\bf r}) \cdot \nabla_p 
\delta Q (t,{\bf r},{\bf p})\rangle &=& 
g \int_0^t dt' \:
\nabla_p^i \langle E^i(t, {\bf r}) E^j\big(t', {\bf r}-{\bf v} (t-t')\big) \rangle \nabla_p^j n({\bf p}) 
+ \nabla_p^i \langle E^i(t, {\bf r}) 
\delta Q_0({\bf r}-{\bf v} t,{\bf p})\rangle \;,
\\ 
\label{E-nabla-dQ}
\langle {\bf E}(t, {\bf r}) \cdot \nabla_p 
\delta \bar Q (t,{\bf r},{\bf p})\rangle &=& 
- g \int_0^t dt' \:
\nabla_p^i \langle E^i(t, {\bf r}) E^j\big(t', {\bf r}-{\bf v} (t-t')\big) \rangle \nabla_p^j \bar n({\bf p}) 
+ \nabla_p^i \langle E^i(t, {\bf r}) 
\delta \bar Q_0({\bf r}-{\bf v} t,{\bf p})\rangle \;,
\\
\nonumber
\langle {\bf E}(t, {\bf r}) \cdot \nabla_p 
\delta G (t,{\bf r},{\bf p})\rangle &=&
g \int_0^t dt' \:
\nabla_p^i \langle E^i(t, {\bf r}) E^j\big(t', {\bf r}-{\bf v} (t-t')\big) \rangle \nabla_p^j n_g({\bf p}) 
+ \nabla_p^i \langle E^i(t, {\bf r}) 
\delta G_0({\bf r}-{\bf v} t,{\bf p})\rangle \;,
\ea

We conclude that the transport equations (\ref{trans-eq-reg}) are determined 
by the correlation functions $\langle E^i(t,{\bf r}) E^j\big(t',{\bf r}')\rangle$, 
$\langle E^i(t, {\bf r}) \delta Q_0({\bf r}',{\bf p})\rangle$,
$\langle E^i(t, {\bf r}) \delta \bar Q_0({\bf r}',{\bf p})\rangle$,
and $\langle E^i(t, {\bf r}) \delta G_0({\bf r}',{\bf p})\rangle$.
To compute these functions, the state of the plasma must be specified.
Although we are mainly interested in an anisotropic plasma, we start with
the isotropic case. Thereafter, we consider the two-stream system.


\section{Isotropic plasma}
\label{sec-iso}

For the case of isotropic plasma, the correlation functions of both 
longitudinal and transverse fields are well known \cite{Mrowczynski:2008ae}. 
Here we limit our considerations to longitudinal chromoelectric 
fields, whose correlation function is \cite{Mrowczynski:2008ae}:
\ba
\label{EE-iso-x}
\langle E^i_a(t,{\bf r}) 
E^j_b(t',{\bf r}') \rangle 
&=& 
\frac{g^2}{2} \, \delta^{ab}
\int_{-\infty +i\sigma}^{\infty +i\sigma}
{d\omega \over 2\pi} \int {d^3k \over (2\pi)^3}
\int_{-\infty +i\sigma'}^{\infty +i\sigma'}
{d\omega' \over 2\pi} \int {d^3k' \over (2\pi)^3} 
e^{-i(\omega t + \omega' t' - 
{\bf k} \cdot {\bf r} - {\bf k}' \cdot {\bf r}')}
\\[2mm] \nonumber
&\times&
\frac{k^i {k'}^j}{{\bf k}^2{{\bf k}'}^2} \:
\frac{(2\pi)^3 \delta^{(3)}({\bf k}' + {\bf k})}
{\varepsilon_L(\omega,{\bf k})\,
\varepsilon_L(\omega',{\bf k}')}
\int {d^3p \over (2\pi)^3} 
\frac{f({\bf p})}
{(\omega - {\bf k} \cdot {\bf v})
(\omega' - {\bf k}' \cdot {\bf v})} 
\;,
\ea
\end{widetext}
where $f({\bf p})\equiv n({\bf p}) + \bar n({\bf p}) + 2N_c n_g({\bf p})$
and $\varepsilon_L(\omega,{\bf k})$ is the longitudinal 
chromodielectric function discussed in the Appendix. 
Note that we do not assume that $n({\bf p})$, $\bar n({\bf p})$,
and $n_g({\bf p})$ are given by the thermal equilibrium distributions,
only that they are isotropic functions of ${\bf p}$. Zeroes of 
$\varepsilon_L(\omega,{\bf k})$ and 
$\varepsilon_L(\omega',{\bf k}')$ as well as those of the 
denominators $(\omega - {\bf k} \cdot {\bf v})$ and
$(\omega' - {\bf k}' \cdot {\bf v})$
contribute to the integrals over $\omega$ and 
$\omega'$. However, when the plasma system under consideration 
is stable with respect to longitudinal modes, all zeroes of 
$\varepsilon_L$ lie in the lower half-plane of complex $\omega$.
Consequently, the contributions associated with these zeroes
exponentially decay in time, and they vanish in the long time 
limit of both $t$ and $t'$. 

We are further interested in the long-time limit of 
$\langle E^i(t,{\bf r}) E^i(t',{\bf r}') \rangle$. The only non-vanishing 
contribution corresponds to the 
poles at $\omega = {\bf k} \cdot {\bf v}$ and 
$\omega' = {\bf k}' \cdot {\bf v}$. This contribution reads
\begin{widetext}
\ba
\label{EE-iso-x-long}
\langle E^i_a(t,{\bf r}) E^j_b(t',{\bf r}') \rangle 
= \frac{g^2}{2} \, \delta^{ab}
\int {d^3p \over (2\pi)^3}
\int {d^3k \over (2\pi)^3}
e^{-i{\bf k} \cdot \big[{\bf v} (t - t') - 
({\bf r} - {\bf r}')\big]}
\frac{k^i k^j}{{\bf k}^4} \:
\frac{ f({\bf p})}{|\varepsilon_L({\bf k} \cdot {\bf v},{\bf k})|^2}
\;,
\ea

The correlation functions like $\langle E^i_a(t,{\bf r}) 
\delta Q_0({\bf r}',{\bf p}') \rangle$ are not computed in 
ref.~\cite{Mrowczynski:2008ae} but they can be readily inferred from
the formulas given there.  One finds
\ba
\nonumber
\langle E^i_a(t,{\bf r}) \delta Q_0({\bf r}',{\bf p}') \rangle
&=& - g \, \tau^a
\int_{-\infty +i\sigma}^{\infty +i\sigma}
{d\omega \over 2\pi} \int {d^3k \over (2\pi)^3}
\int {d^3k' \over (2\pi)^3}
e^{-i(\omega t - {\bf k} \cdot {\bf r}
- {\bf k}' \cdot {\bf r}')}
\frac{k^i}{{\bf k}^2} \:
\frac{(2\pi )^3 \delta^{(3)}({\bf k}' + {\bf k})}
{\varepsilon_L(\omega,{\bf k})} \:
\frac{n({\bf p}')}
{\omega - {\bf k}\cdot {\bf v}'} 
\\[2mm] 
\label{corr-E-dQ-iso-x}
&=& i g \, \tau^a 
\int {d^3k \over (2\pi)^3}
e^{-i{\bf k}\cdot ({\bf v}' t - {\bf r} + {\bf r}'))}
\frac{k^i}{{\bf k}^2} \:
\frac{ n({\bf p}')}
{\varepsilon_L({\bf k}\cdot {\bf v}',{\bf k})} 
\;,
\ea
where the last equality holds in the long-time limit which is carried 
by the contribution corresponding to the pole 
$\omega = {\bf k} \cdot {\bf v}'$. Similarly, one finds
\ba
\nonumber
\langle E^i_a(t,{\bf r}) \delta \bar Q_0({\bf r}',{\bf p}') \rangle
&=& g \, \tau^a
\int_{-\infty +i\sigma}^{\infty +i\sigma}
{d\omega \over 2\pi} \int {d^3k \over (2\pi)^3}
\int {d^3k' \over (2\pi)^3} 
e^{-i(\omega t - {\bf k} \cdot {\bf r} 
- {\bf k}' \cdot {\bf r}')}
\frac{k^i}{{\bf k}^2} \:
\frac{(2\pi )^3 \delta^{(3)}({\bf k}' + {\bf k})}
{\varepsilon_L(\omega,{\bf k})} \:
\frac{\bar n({\bf p}')}
{\omega - {\bf k}\cdot {\bf v}'} 
\\[2mm] 
\label{corr-E-dbarQ-iso-x}
&=& - i g \, \tau^a 
\int {d^3k \over (2\pi)^3}
e^{-i{\bf k}\cdot ({\bf v}' t - {\bf r} + {\bf r}'))}
\frac{k^i}{{\bf k}^2} \:
\frac{\bar n({\bf p}')}
{\varepsilon_L({\bf k}\cdot {\bf v}',{\bf k})} 
\;,
\ea
\ba
\nonumber
\langle E^i_a(t,{\bf r}) \delta G_0({\bf r}',{\bf p}') \rangle
&=& - g \, T^a
\int_{-\infty +i\sigma}^{\infty +i\sigma}
{d\omega \over 2\pi} \int {d^3k \over (2\pi)^3}
\int {d^3k' \over (2\pi)^3} 
e^{-i(\omega t - {\bf k} \cdot {\bf r} 
- {\bf k}' \cdot {\bf r}')}
\frac{k^i}{{\bf k}^2} \:
\frac{(2\pi )^3 \delta^{(3)}({\bf k}' + {\bf k})}
{\varepsilon_L(\omega,{\bf k})} \:
\frac{n_g({\bf p}')}
{\omega - {\bf k}\cdot {\bf v}'} 
\\[2mm] 
\label{corr-E-dG-iso-x}
&=& i g \, T^a 
\int {d^3k \over (2\pi)^3}
e^{-i{\bf k}\cdot ({\bf v}' t - {\bf r} + {\bf r}'))}
\frac{k^i}{{\bf k}^2} \:
\frac{ n_g({\bf p}')}
{\varepsilon_L({\bf k}\cdot {\bf v}',{\bf k})} 
\;.
\ea

Substituting the correlation functions (\ref{EE-iso-x-long},
\ref{corr-E-dQ-iso-x}) into (\ref{E-nabla-dQ}), one finds 
\ba
\nonumber
{\rm Tr}\langle {\bf E}(t, {\bf r}) \cdot \nabla_p 
\delta Q(t,{\bf r},{\bf p})\rangle 
&=&
\frac{g^3}{4}(N_c^2-1)
\int_0^t dt'
\nabla_p^i 
\int {d^3p' \over (2\pi)^3}
\int {d^3k \over (2\pi)^3}
e^{i{\bf k} \cdot ({\bf v}- {\bf v}') (t - t')}
\frac{k^i k^j}{{\bf k}^4} \:
\frac{ f({\bf p}')}{|\varepsilon_L({\bf k} \cdot {\bf v}',{\bf k})|^2}
\; \nabla_p^j n({\bf p}) 
\\[2mm]
\label{E-nabla-dQ-2}
&+&
i \frac{g}{2}(N_c^2-1) \nabla_p^i
\int {d^3k \over (2\pi)^3}
\frac{k^i}{{\bf k}^2} \:
\frac{ n({\bf p})}
{\varepsilon_L({\bf k}\cdot {\bf v},{\bf k})}
\;,
\ea
and analogous expressions for 
${\rm Tr}\langle {\bf E}(t, {\bf r}) \cdot \nabla_p
\delta \bar Q(t,{\bf r},{\bf p})\rangle$ and
${\rm Tr}\langle {\bf E}(t, {\bf r}) \cdot \nabla_p
\delta G(t,{\bf r},{\bf p})\rangle$.
As shown in \cite{Mrowczynski:2008ae},
${\rm Tr}\langle E^i(t,{\bf r}) E^j(t',{\bf r}') \rangle$
is gauge independent within the linear response approach. 
The same arguments used to show this apply to 
${\rm Tr}\langle E^i(t,{\bf r}) \delta Q_0({\bf r}',{\bf p}')\rangle$.
Thus, we conclude that the collision
term of the transport equation (\ref{trans-eq-reg}),
${\rm Tr} \langle {\bf E}(t, {\bf r}) \cdot \nabla_p 
\delta Q (t,{\bf r},{\bf p})\rangle$, is gauge independent.

Let us now discuss the first term on the r.h.s.\ of (\ref{E-nabla-dQ-2}).
Computing the integral over $t'$ we get
\ba 
{\rm Tr}\langle {\bf E}(t, {\bf r}) \cdot \nabla_p
\delta Q(t,{\bf r},{\bf p})\rangle_{(1)}
&=&
\frac{g^3}{4}(N_c^2-1) \:
\nabla_p^i
\int {d^3p' \over (2\pi)^3}
\int {d^3k \over (2\pi)^3} \frac{k^i k^j}{{\bf k}^4} \:
\frac{ f({\bf p}')}{|\varepsilon_L({\bf k} \cdot {\bf v}',{\bf k})|^2}
\\[2mm] \nonumber
&\times&
\Big( -i
\frac{\cos ({\bf k} \cdot ({\bf v}- {\bf v}')t) - 1}
{ {\bf k} \cdot ({\bf v}- {\bf v}')}
+
\frac{\sin ({\bf k} \cdot ({\bf v}- {\bf v}')t)}
{{\bf k} \cdot ({\bf v}- {\bf v}')} \Big)
\; \nabla_p^j n({\bf p}) \;.
\ea
The first term does not contribute to the integral because it is an odd
function of ${\bf k}$. Since in the limit $t \rightarrow \infty$ we have
\be
\lim_{t \rightarrow \infty}
\frac{\sin ({\bf k} \cdot ({\bf v}- {\bf v}')t)}
{{\bf k} \cdot ({\bf v}- {\bf v}')}
= \pi \delta\big({\bf k} \cdot ({\bf v}- {\bf v}')\big)\;,
\ee
one finally finds
\ba
\label{E-nabla-dQ-1-final}
{\rm Tr}\langle {\bf E}(t, {\bf r}) \cdot \nabla_p
\delta Q(t,{\bf r},{\bf p})\rangle_{(1)}
= \frac{g^3}{4} \,\pi \, (N_c^2-1) \:
\nabla_p^i
\int {d^3p' \over (2\pi)^3}
\int {d^3k \over (2\pi)^3} \frac{k^i k^j}{{\bf k}^4} \:
\frac{ f({\bf p}')}{|\varepsilon_L({\bf k} \cdot {\bf v}',{\bf k})|^2} \:
\delta \big({\bf k} \cdot ({\bf v}- {\bf v}')\big) \; \nabla_p^j n({\bf p})
\,.
\ea
Analogously, one computes 
${\rm Tr}\langle {\bf E}(t, {\bf r}) \cdot \nabla_p
\delta \bar Q(t,{\bf r},{\bf p})\rangle_{(1)}$ and
${\rm Tr}\langle {\bf E}(t, {\bf r}) \cdot \nabla_p
\delta G(t,{\bf r},{\bf p})\rangle_{(1)}$.

The second term on the r.h.s.\ of (\ref{E-nabla-dQ-2}) can be written as
\ba
\label{E-nabla-dQ0}
{\rm Tr}\langle {\bf E}(t,{\bf r}) \cdot \nabla_p 
\delta Q(t,{\bf r},{\bf p})\rangle_{(2)} =
\frac{g}{2}(N_c^2-1) \nabla_p^i
\int {d^3k \over (2\pi)^3}
\frac{k^i}{{\bf k}^2} \:
\frac{ \Im \varepsilon_L({\bf k}\cdot {\bf v},{\bf k})}
{|\varepsilon_L({\bf k}\cdot {\bf v},{\bf k})|^2} 
\; n({\bf p})
\;,
\ea
because the term with $\Re \varepsilon_L({\bf k}\cdot {\bf v},{\bf k})$
is an odd function of ${\bf k}$ (see Appendix). Alternatively, one can argue 
that the r.h.s.\ of (\ref{E-nabla-dQ0}) has to be real as the l.h.s. is real. 
In the same way one finds:
$\langle {\bf E}(t, {\bf r}) \cdot \nabla_p
\delta \bar Q(t,{\bf r},{\bf p})\rangle_{(2)}$ and
$\langle {\bf E}(t, {\bf r}) \cdot \nabla_p
\delta G(t,{\bf r},{\bf p})\rangle_{(2)}$. 

With the formulas derived above, the transport equations (\ref{trans-eq-reg}) 
can now be written either in the Balescu-Lenard form or the Fokker-Planck form.
\end{widetext}

\subsection{Balescu-Lenard equations}

Using the formula (\ref{Im-eL}) to express $\Im \varepsilon_L$ through
the distribution function, the transport equations (\ref{trans-eq-reg})
get the Balescu-Lenard form \cite{LP81}
\ba
\nonumber 
{\cal D}\,  n(t,{\bf r},{\bf p})
&=& \nabla_p \cdot {\bf S}[n,\bar n,n_g] \;, 
\\  
\label{BL-eq-iso}
{\cal D}\, \bar n(t,{\bf r},{\bf p}) 
&=& 
\nabla_p \cdot \bar{\bf S}[n,\bar n,n_g] \;, 
\\ \nonumber 
{\cal D}\,  n_g(t,{\bf r},{\bf p})
&=&  \nabla_p \cdot {\bf S}_g[n,\bar n,n_g] \;,
\ea
where, as previously, ${\cal D}$ is the material derivative, and 
\ba
\nonumber 
S^i[n,\bar n,n_g] 
&\equiv& 
\int {d^3p' \over (2\pi)^3}
B^{ij}({\bf v},{\bf v}') 
\\ \nonumber 
&\times&
\Big[\nabla_p^j n({\bf p}) \, f({\bf p}') 
- n({\bf p}) \, \nabla_{p'}^j f({\bf p}')\Big] \,,
\\ \label{coll-terms-BL} 
\bar S^i[n,\bar n,n_g] 
&\equiv& 
\int {d^3p' \over (2\pi)^3}
B^{ij}({\bf v},{\bf v}')
\\ \nonumber 
&\times&
\Big[\nabla_p^j \bar n({\bf p}) \, f({\bf p}') 
- \bar n({\bf p}) \, \nabla_{p'}^j f({\bf p}')\Big] \,,
\\  \nonumber 
S^i_g[n,\bar n,n_g] 
&\equiv& 
\int {d^3p' \over (2\pi)^3}
B^{ij}_g({\bf v},{\bf v}')
\\ \nonumber 
&\times&
\Big[\nabla_p^j n_g({\bf p}) \, f({\bf p}') 
- n_g({\bf p}) \, \nabla_{p'}^j f({\bf p}')\Big] \,,
\ea
with 
\be
B^{ij}({\bf v},{\bf v}') \equiv \frac{g^4}{8}
\frac{N_c^2-1}{N_c} 
\int {d^3k \over (2\pi)^3} \frac{k^i k^j}{{\bf k}^4} \:
\frac{2\pi\delta \big({\bf k} \cdot ({\bf v}-{\bf v}')\big)}
{|\varepsilon_L({\bf k} \cdot {\bf v},{\bf k})|^2},
\ee
and 
\be
B^{ij}_g({\bf v},{\bf v}') = \frac{2N_c^2}{N_c^2-1} 
B^{ij}({\bf v},{\bf v}') \,.
\ee
Since the interaction processes that are taken into account 
conserve the numbers of particles of every species ($q,\bar{q},g$), 
the transport equations in the Balescu-Lenard form 
(\ref{BL-eq-iso}) can be seen as continuity equations in 
momentum space with ${\bf S}$, $\bar{\bf S}$, ${\bf S}_g$
playing a role of currents. 
One observes that for classical equilibrium functions 
\be
f^{\rm eq} ({\bf p}), n^{\rm eq} ({\bf p}), 
\bar n^{\rm eq} ({\bf p}), n^{\rm eq}_g ({\bf p}) 
\sim e^{-E_p/T} \; , 
\ee
the collision terms (\ref{coll-terms-BL}) vanish, as expected, 
because 
\be
(v^i - {v'}^i) B^{ij}({\bf v},{\bf v}') = 0 \; .
\ee

If $\varepsilon_L(\omega,{\bf k})$ is replaced by unity, {\it i.e.} if
one ignores the chromodielectric properties of the plasma, 
the tensor $B^{ij}({\bf v},{\bf v}')$ is easily found to be
\ba
\label{B-Lan}
B^{ij}({\bf v},{\bf v}') &=& \frac{g^4}{32 \pi}\frac{N_c^2-1}{N_c} \,
\frac{L}{|{\bf v}-{\bf v}'|} 
\\ \nonumber 
&& \times \bigg( \delta^{ij}
- \frac{(v^i - {v'}^i) (v^j - {v'}^j)}{({\bf v}-{\bf v}')^2}  \bigg)\;,
\ea
with 
\begin{equation}
L \equiv \int dk/k = \ln (k_{\rm max}/ k_{\rm min}) \;.
\end{equation}
The parameter $L$ is called the Coulomb logarithm and the collision
term with the tensor $B^{ij}({\bf v},{\bf v}')$ of the form 
(\ref{B-Lan}) is called the Landau collision term \cite{LP81}. 
Estimating $k_{\rm max}$ as the system temperature $T$ and 
$k_{\rm min}$ as the Debye mass $m_D \sim gT$, one finds 
$L \sim \ln (1/g)$. 

It may appear strange that we start with the collisionless 
transport equations (\ref{transport-eq}) to derive the collision 
terms. This procedure, which is commonly used in the 
plasma literature, is well justified, however, see {\it e.g.} 
Ref.~\cite{LP81}. The collision terms, which are derived above, 
represent the effect of fluctuating soft fields on the hard 
quasiparticles. It is important to note that the collision terms 
are dominated, as it should be, by the soft wave vectors. 
Consequently, the collisions of quasiparticles involving the 
exchange of hard momenta, which are neglected in 
Eqs.~(\ref{transport-eq}), do not need to be taken into 
account at lowest order.

\subsection{Fokker-Planck equations}

Sometimes it is more convenient to use the transport equations 
in the Fokker-Planck form. Following Ref.~\cite{LP81},
one rewrites Eqs.~(\ref{BL-eq-iso}) as
\ba
\nonumber 
\big( {\cal D}  
- \nabla_p^i X^{ij}({\bf v}) \nabla_p^j 
- \nabla_p^i Y^i({\bf v}) \big)
n(t,{\bf r},{\bf p})
&=& 0 \;, 
\\ 
\label{FP-eq-iso}
\big( {\cal D}  
- \nabla_p^i  X^{ij}({\bf v}) \nabla_p^j 
- \nabla_p^i  Y^i({\bf v}) \big)
\bar n(t,{\bf r},{\bf p}) 
&=& 0 \;, 
\\ \nonumber 
\big( {\cal D}  
- \nabla_p^i X^{ij}_g({\bf v}) \nabla_p^j 
- \nabla_p^i Y^i_g({\bf v}) \big)
n_g(t,{\bf r},{\bf p})
&=&  0\;,
\ea
where
\begin{widetext}
\ba
X^{ij}({\bf v}) &\equiv&
\frac{g^4}{8}(N_c^2-1)
\int {d^3p' \over (2\pi)^3}
\int {d^3k \over (2\pi)^3} \frac{k^i k^j}{{\bf k}^4} \:
\frac{ f({\bf p}')}{|\varepsilon_L({\bf k} \cdot {\bf v}',{\bf k})|^2} \:
2\pi \delta \big({\bf k} \cdot ({\bf v}- {\bf v}')\big)
=
\int {d^3p' \over (2\pi)^3} f({\bf p}') \, B^{ij}({\bf v},{\bf v}') \;,
\\ [2mm]
Y^i({\bf v}) &\equiv&
\frac{g^2}{2}(N_c^2-1) 
\int {d^3k \over (2\pi)^3}
\frac{k^i}{{\bf k}^2} \:
\frac{ \Im \varepsilon_L({\bf k}\cdot {\bf v},{\bf k})}
{|\varepsilon_L({\bf k}\cdot {\bf v},{\bf k})|^2} 
\\[2mm] \nonumber
&=&
-\frac{g^4}{8}(N_c^2-1) 
\int {d^3p' \over (2\pi)^3} 
\int {d^3k \over (2\pi)^3}
\frac{k^i}{{\bf k}^4} \:
\frac{{\bf k} \cdot \nabla_{p'} f({\bf p}')}
{|\varepsilon_L({\bf k}\cdot {\bf v},{\bf k})|^2} 
\, 2\pi \delta  \big({\bf k} \cdot ({\bf v}- {\bf v}')\big)
=
- \int {d^3p' \over (2\pi)^3} 
\nabla_{p'}^j f({\bf p}') \, B^{ij}({\bf v},{\bf v}')
\;,
\ea
\end{widetext}
and 
\begin{eqnarray}
X^{ij}_g({\bf v}) &= & \frac{2N_c^2}{N_c^2-1} X^{ij}({\bf v}) \;,
\\
Y^i_g({\bf v}) &= & \frac{2N_c^2}{N_c^2-1} Y^i({\bf v}) \;. 
\end{eqnarray}
The equations (\ref{FP-eq-iso}) appear to be linear but actually 
they are not: the coefficients  $X^{ij}({\bf v})$, $Y^i({\bf v})$, 
$X^{ij}_g({\bf v})$ and $Y^i_g({\bf v})$ depend on the distribution 
functions. When the distribution functions are of the classical 
equilibrium form ($f^{\rm eq} ({\bf p})$, $n^{\rm eq} ({\bf p})$, 
$\bar n^{\rm eq} ({\bf p})$, $n^{\rm eq}_g ({\bf p}) \sim e^{-E_p/T}$), 
we have the relation
\be
\label{relation-eq1}
Y^i({\bf v}) = \frac{v^i}{T} X^{ij}({\bf v}) \;.
\ee
Consequently, the Fokker-Planck collision terms vanish in 
equilibrium, as do the Balescu-Lenard collision terms. 

Since the system is assumed to be isotropic, $X^{ij}({\bf v})$ and 
$Y^i({\bf v})$ can be expressed as follows:
\ba
\label{Xij}
X^{ij}({\bf v}) &=& a \, \delta^{ij} + b \, v^i v^j \;,
\\ [2mm]
\label{Yi}
Y^i ({\bf v}) &=& c \, v^i \;,
\ea
with
\ba
\label{form-a}
a  &=&
\frac{1}{2} \int {d^3p' \over (2\pi)^3} f({\bf p}') \, 
\Big[\delta^{ji} - v^j v^i \Big] B^{ij}({\bf v},{\bf v}')
\\ [2mm]
\label{form-b}
b &=&
\frac{1}{2} \int {d^3p' \over (2\pi)^3} f({\bf p}') \, 
\Big[3 v^j v^i - \delta^{ji}\Big] B^{ij}({\bf v},{\bf v}')
\\ [2mm]
\label{form-c}
c &=&
- \int {d^3p' \over (2\pi)^3} v^i
\nabla_{p'}^j f({\bf p}') \, B^{ij}({\bf v},{\bf v}')
\;.
\ea
Because of the system's isotropy, the coefficients $a$, $b$, $c$
can depend only on ${\bf v}^2$. In the ultrarelativistic
limit, which is adopted here, ${\bf v}^2 = 1$, and consequently 
$a$, $b$, $c$ are independent of ${\bf v}$. We also note that
in equilibrium the coefficients are related as
\be
\label{relation-eq2}
T \, c = a + b  \;,
\ee
which follows from Eq.~(\ref{relation-eq1}).

When $\varepsilon_L(\omega,{\bf k})$ is replaced, as previously, 
by unity one finds that $b=0$ and 
\ba
\label{form-a-1}
a &\equiv& \frac{g^4}{96 \pi^3}(N_c^2-1) \, L \,
\int_0^\infty dp \, p^2 f({\bf p}) \;,
\\[2mm]
c &\equiv& -\frac{g^4}{96 \pi^3}(N_c^2-1) \, L \,
\int_0^\infty dp \, p^2 \frac{df({\bf p})}{dp} \;.
\ea
Using the relations (\ref{m_D^2}), the coefficient $c$ can be expressed 
in terms of the Debye mass as
\be
c = \frac{g^2}{24 \pi^2}(N_c^2-1) \, L \,m_D^2.
\ee
Furthermore, in equilibrium, $a=c \,T$.

We note that in spite of our neglect of transverse chromodynamic fields, 
the collision terms for the isotropic plasma derived here are very similar to 
those derived in \cite{Selikhov:1991br,Selikhov:1994xn,Bodeker:1999ey,Arnold:1998cy,Litim:1999ns,Litim:1999id,ValleBasagoiti:1999pg,Blaizot:1999xk,Markov:2000pb}.


\subsection{Equilibration of an isotropic plasma}


As an application of the Fokker-Planck equations (\ref{FP-eq-iso})
we discuss the problem of plasma equilibration. In this section we limit
our considerations to quarks, as the analysis for antiquarks and gluons
is very similar. We consider the system which is homogenous and mostly 
equilibrated but a small fraction $(\lambda \ll 1)$ of the particles, denoted 
by $\delta n(t,{\bf p})$, is out of equilibrium. One asks on what time scale 
the system reaches the equilibrium. The distribution 
function  is assumed to be of the form
\be
n(t,{\bf p}) = (1-\lambda) \, n^{\rm eq}({\bf p}) 
+ \lambda \,\delta n(t,{\bf p}) \;.
\ee
In the course of equilibration $n(t,{\bf p})$ tends to 
$n^{\rm eq}({\bf p})$. Since the particle number is conserved within 
the transport theory approach developed here, $\delta n(t,{\bf p})$ is 
not reduced to zero in the equilibration process but it tends to 
$n^{\rm eq}({\bf p})$.

We define the rate of equilibration $\Gamma $ through the relation
\be 
\label{rate-eq-def}
\frac{\partial n}{\partial t} = \Gamma \delta n \;.
\ee
We note that $\Gamma$ is either positive, when $\delta n$ grows
going to $n^{\rm eq}$, and it is negative, when $\delta n$ 
decreases going to $n^{\rm eq}$. Using the Fokker-Planck equation 
(\ref{FP-eq-iso}), the definition (\ref{rate-eq-def}) gives
\be
\label{rate-eq-1}
\Gamma =  \frac{1}{\delta n}
\big(\nabla_p^i X^{ij} \nabla_p^j + \nabla_p^i Y^i \big)
\delta n \;.
\ee

Since the fraction of particles with non-equilibrium distribution
is assumed to be small, the coefficients $a$, $b$, $c$ from the 
formulas (\ref{Xij}, \ref{Yi}) are given by the equilibrium 
function $n^{\rm eq} \sim e^{- E_p/T}$. Using the approximate expression 
of $a$ (\ref{form-a-1}) with $b=0$ and $c = a/T$, Eq.~(\ref{rate-eq-1})
is rewritten  as
\be
\label{rate-eq-2}
\Gamma =  \frac{a}{\delta n}
\Big(\nabla_p^2 + \frac{1}{T}{\bf v} \cdot \nabla_p 
+ \frac{2}{T E_p}\Big) \delta n \;.
\ee
\smallskip

The equilibration rate obviously depends of the form of $\delta n$.
Here we consider the case where the small fraction of partons has an 
equilibrium distribution of temperature $T_0$ which differs from the 
temperature $T$ of the bulk of the partons. Thus, 
$\delta n \sim e^{- E_p/T_0}$. Then, the equilibration rate 
(\ref{rate-eq-2}) equals
\be
\label{rate-eq-final}
\Gamma =  a\, \frac{T-T_0}{T_0^2 T \, E_p} \,
(E_p - 2T_0) \;.
\ee
For $T = T_0$, the whole system is in equilibrium and, as
expected, $\Gamma = 0$. When $T > T_0$, the distribution 
$e^{- E_p/T_0}$ is steeper than $e^{- E_p/T}$. 
Equation (\ref{rate-eq-final}) tells us that $\delta n$ decreases 
for $E_p < 2T_0$ and grows for $E_p > 2T_0$ during the 
equilibration process. When $T < T_0$, we have the opposite situation. 
In both cases, the slope of the distribution function $\delta n$ 
tends to the slope of $n^{\rm eq}$. With the coefficient $a$ 
given by (\ref{form-a-1}), the formula (\ref{rate-eq-final}) 
quantitatively predicts how fast the equilibrium is approached.


\section{Two-stream system}
\label{sec-2-streams}


The two-stream configuration provides an interesting case of
an unstable plasma. The correlation function of longitudinal 
chromoelectric fields, which is needed to derive the transport 
equations, was computed in \cite{Mrowczynski:2008ae}. Unfortunately
the correlation function for transverse fields is not known. 
This limits our considerations to longitudinal fields. 

The distribution function of the two-stream system is chosen as
\be
\label{f-2-streams}
f({\bf p}) = (2\pi )^3 n 
\Big[\delta^{(3)}({\bf p} - {\bf q}) 
+ \delta^{(3)}({\bf p} + {\bf q}) \Big] \;,
\ee
where $n$ is the effective parton density in a single stream. The 
distribution function (\ref{f-2-streams}) should be treated as an 
idealization of the two-peak distribution where the particles have 
momenta close to ${\bf q}$ or $-{\bf q}$. 

To compute $\varepsilon_L(\omega,{\bf k})$ we first perform an integration 
by parts in (\ref{eL}) and then substitute the distribution function 
(\ref{f-2-streams}) into the resulting formula. We obtain 
\begin{widetext}
\ba
\label{eL-3}
\varepsilon_L(\omega,{\bf k}) 
= 1 - \mu^2 
\frac{{\bf k}^2 -({\bf k} \cdot {\bf u})^2}{{\bf k}^2}
\bigg[
  \frac{1}{(\omega - {\bf k} \cdot {\bf u})^2}
+ \frac{1}{(\omega + {\bf k} \cdot {\bf u})^2}
\bigg]
\\[2mm] \nonumber
= \frac{\big(\omega - \omega_+({\bf k})\big)
\big(\omega + \omega_+({\bf k})\big)
\big(\omega - \omega_-({\bf k})\big)
\big(\omega + \omega_-({\bf k})\big)}
{\big(\omega^2 - ({\bf k} \cdot {\bf u})^2\big)^2} 
\;,
\ea
where ${\bf u} \equiv {\bf q}/E_{\bf q}$ is the stream velocity, 
$\mu^2 \equiv g^2n/2 E_{\bf q}$ and $\pm \omega_{\pm}({\bf k})$
are the four roots of the dispersion equation 
$\varepsilon_L(\omega,{\bf k}) = 0$ which are explicitly given by
\be
\label{roots}
\omega_{\pm}^2({\bf k}) = \frac{1}{{\bf k}^2}
\bigg[{\bf k}^2 ({\bf k} \cdot {\bf u})^2
+ \mu^2 \big({\bf k}^2 - ({\bf k} \cdot {\bf u})^2\big)
\pm \mu \sqrt{\big({\bf k}^2 - ({\bf k} \cdot {\bf u})^2\big)
\Big(4{\bf k}^2 ({\bf k} \cdot {\bf u})^2 +
\mu^2 \big({\bf k}^2 - ({\bf k} \cdot {\bf u})^2\big)\Big)} 
\; \bigg] \;.
\ee
One can show that $0 < \omega_+({\bf k}) \in \mathbb{R}$ for any 
${\bf k}$, while $\omega_-({\bf k})$ is imaginary for 
${\bf k} \cdot {\bf u} \not=0$ and 
${\bf k}^2 ({\bf k} \cdot {\bf u})^2 
< 2 \mu^2 \big({\bf k}^2 - ({\bf k} \cdot {\bf u})^2\big)$.
$\omega_-$ represents the well-known two-stream electrostatic 
instability generated by a mechanism analogous to
the Landau damping. For ${\bf k}^2 ({\bf k} \cdot {\bf u})^2 
\ge 2 \mu^2 \big({\bf k}^2 - ({\bf k} \cdot {\bf u})^2\big)$, 
the $\omega_-$ mode is stable: $0 < \omega_-({\bf k}) \in \mathbb{R}$. 

The terms like $\langle {\bf E}(t, {\bf r}) \cdot \nabla_p 
\delta Q (t,{\bf r},{\bf p})\rangle$, which enter the transport 
equations (\ref{trans-eq-reg}) are given by Eqs.~(\ref{E-nabla-dQ}).
As for the isotropic plasma one needs to specify the
the correlation functions $\langle E_a^i(\omega,{\bf k}) 
E_b^i(\omega',{\bf k}') \rangle$,
$\langle {\bf E}(t,{\bf r}) \delta Q_0 ({\bf r}',{\bf p}')\rangle$,
etc.  The correlation function of the longitudinal fields
$\langle E_a^i(\omega,{\bf k}) E_b^j(\omega',{\bf k}') \rangle$ 
was found in \cite{Mrowczynski:2008ae}:
\ba
\label{E-fluc-2-stream-2} 
\langle E_a^i(\omega,{\bf k}) 
E_b^j(\omega',{\bf k}') \rangle 
&=& - g^2 \delta^{ab} n \: 
\frac{(2\pi)^3\delta^{(3)}({\bf k} + {\bf k}')}
{{\bf k}^2} 
\frac{k^i k^j} {{\bf k}^2}
\Big[\omega \omega' + 
({\bf k} \cdot {\bf u})({\bf k}' \cdot {\bf u}) \Big]
\\ [2mm] \nonumber 
&\times&
\frac{\omega^2 - ({\bf k} \cdot {\bf u})^2} 
{\big(\omega - \omega_-({\bf k})\big)
\big(\omega + \omega_-({\bf k})\big)
\big(\omega - \omega_+({\bf k})\big)
\big(\omega + \omega_+({\bf k})\big)}
\\ [2mm] \nonumber 
&\times&
\frac{{\omega'}^2 - ({\bf k}' \cdot {\bf u})^2} 
{\big(\omega' - \omega_-({\bf k}')\big)
\big(\omega' + \omega_-({\bf k}')\big)
\big(\omega' - \omega_+({\bf k}')\big)
\big(\omega' + \omega_+({\bf k}')\big) }
\;.
\ea

We are particularly interested in the contributions of the unstable modes
to the correlation function. For this reason we consider the domain of wave 
vectors obeying ${\bf k} \cdot {\bf u} \not=0$ and 
${\bf k}^2 ({\bf k} \cdot {\bf u})^2 < 2 \mu^2 \big({\bf k}^2 - ({\bf k} \cdot {\bf u})^2\big)$
when $\omega_-({\bf k})$ is imaginary and the mode is unstable. We write 
$\omega_-({\bf k})=i \gamma_{\bf k}$ 
with $0 < \gamma_{\bf k} \in \mathbb{R}$. The contribution coming from the 
modes $\pm \omega_-({\bf k})$ then equals \cite{Mrowczynski:2008ae} 
\ba
\label{EL-fluc-x-stream7}
\langle E_a^i(t,{\bf r}) E_b^j(t',{\bf r}') 
\rangle_{\rm unstable}
&=& \frac{g^2}{2}\,\delta^{ab} n 
\int {d^3k \over (2\pi)^3} 
\frac{e^{i {\bf k}({\bf r} - {\bf r}')}}{{\bf k}^4}
\frac{k^i k^j}{(\omega_+^2 - \omega_-^2)^2}
\frac{
\big(\gamma_{\bf k}^2 + ({\bf k} \cdot {\bf u})^2\big)^2} 
{\gamma_{\bf k}^2}
\\[2mm] \nonumber
&\times& 
\Big[
\big(\gamma_{\bf k}^2 + ({\bf k} \cdot {\bf u})^2\big)
\cosh \big(\gamma_{\bf k} (t + t')\big)
+
\big(\gamma_{\bf k}^2 - ({\bf k} \cdot {\bf u})^2\big)
\cosh \big(\gamma_{\bf k} (t - t')\big) \Big] \;.
\ea
\end{widetext}
As Eq.~(\ref{EL-fluc-x-stream7}) shows, the contribution of the unstable modes
to the field-field correlation function is space translation invariant -- 
it depends only on the difference $({\bf r} - {\bf r}')$. If the initial plasma 
is on average homogeneous, it remains so over the course of
its evolution. The time dependence of the correlation function 
(\ref{EL-fluc-x-stream7}), however, is very different from the spatial
dependence. The electric field grows exponentially and so does
the correlation function, both in $(t + t')$ and $(t - t')$.
The fluctuation spectrum also evolves in time as the growth rate
of the unstable modes is wave-vector dependent. After a sufficiently
long time the fluctuation spectrum will be dominated by the fastest 
growing modes.

The correlation function
$\langle {\bf E}(t,{\bf r}) \delta Q_0 ({\bf r}-{\bf v}t,{\bf p})\rangle$
is, as previously, given by Eqs.~(\ref{E-nabla-dQ0}). Since the dielectric
function (\ref{eL-3}) is real, the correlation functions 
$\langle {\bf E}(t,{\bf r}) \delta Q_0 ({\bf r}-{\bf v}t,{\bf p})\rangle$,
$\langle {\bf E}(t,{\bf r}) \delta \bar Q_0({\bf r}-{\bf v}t,{\bf p})\rangle$
and $\langle {\bf E}(t,{\bf r}) \delta G_0 ({\bf r}-{\bf v}t,{\bf p})\rangle$
all vanish. Therefore,
\begin{widetext}
\ba
\nonumber 
{\rm Tr}\langle {\bf E}(t, {\bf r}) \cdot \nabla_p 
\delta Q (t,{\bf r},{\bf p})\rangle 
&=& 
g \int_0^t dt' \:
\nabla_p^i \langle E^i(t, {\bf r}) E^j\big(t', {\bf r}-{\bf v} (t-t')\big) \rangle \nabla_p^j n({\bf p}) 
\\ [2mm] 
&=&
\frac{g^3}{4}\,(N_c^2-1) \, n
\int_0^t dt' \: \nabla_p^i 
\int {d^3k \over (2\pi)^3}
\frac{e^{i {\bf k}{\bf v} (t-t')}}{{\bf k}^4}
\frac{k^i k^j }{(\omega_+^2 - \omega_-^2)^2}
\frac{\big(\gamma_{\bf k}^2 + ({\bf k} \cdot {\bf u})^2\big)^2} 
{\gamma_{\bf k}^2}
\\[2mm] \nonumber
&\times& 
\Big[
\big(\gamma_{\bf k}^2 + ({\bf k} \cdot {\bf u})^2\big)
\cosh \big(\gamma_{\bf k} (t + t')\big)
+
\big(\gamma_{\bf k}^2 - ({\bf k} \cdot {\bf u})^2\big)
\cosh \big(\gamma_{\bf k} (t - t')\big) \Big] \;
\nabla_p^j n({\bf p})\;.
\ea
Performing the integration over $t'$ and keeping only the real part,
one finds
\ba
{\rm Tr}\langle {\bf E}(t, {\bf r}) \cdot \nabla_p 
\delta Q (t,{\bf r},{\bf p})\rangle 
&=& 
\frac{g^3}{4} \, (N_c^2-1) \, n \: \nabla_p^i 
\int {d^3k \over (2\pi)^3} 
\frac{k^i k^j }{{\bf k}^4 (\omega_+^2 - \omega_-^2)^2}
\frac{\big(\gamma_{\bf k}^2 + ({\bf k} \cdot {\bf u})^2\big)^2} 
{\gamma_{\bf k}^2 \big(\gamma_{\bf k}^2 + ({\bf k}{\bf v})^2\big)}
\\[2mm] \nonumber
&\times& 
\bigg[\gamma_{\bf k}^3 \sinh(2 \gamma_{\bf k}t)
+ ({\bf k} \cdot {\bf u})^2 
\Big(\gamma_{\bf k} \sinh(2 \gamma_{\bf k}t)
\\[2mm] \nonumber
&+& 
({\bf k} \cdot {\bf v}) \, \sin\big(({\bf k} \cdot {\bf v})t\big) \,
     \cosh(\gamma_{\bf k}t)
- \gamma_{\bf k} \cos\big(({\bf k} \cdot {\bf v})t\big) \,
     \sinh(\gamma_{\bf k}t)
\Big) \bigg]  \nabla_p^j n({\bf p})  \:.
\ea
Neglecting the oscillating terms, we finally get
\ba
{\rm Tr}\langle {\bf E}(t, {\bf r}) \cdot \nabla_p
\delta Q (t,{\bf r},{\bf p})\rangle
&=&
\frac{g^3}{2}\, (N_c^2-1) \, n \: \nabla_p^i 
\int {d^3k \over (2\pi)^3}
\frac{k^i k^j }{{\bf k}^4 (\omega_+^2 - \omega_-^2)^2}
\frac{\big(\gamma_{\bf k}^2 + ({\bf k} \cdot {\bf u})^2\big)^3}
{\gamma_{\bf k} \big(\gamma_{\bf k}^2 + ({\bf k}{\bf v})^2\big)}\,
\sinh(2 \gamma_{\bf k}t) \; \nabla_p^j n({\bf p}) \,.
\ea
In an analogous way one can obtain explicit expressions for
$\langle {\bf E}(t, {\bf r}) \cdot \nabla_p 
\delta \bar Q (t,{\bf r},{\bf p})\rangle$ and
$\langle {\bf E}(t, {\bf r}) \cdot \nabla_p 
\delta G (t,{\bf r},{\bf p})\rangle$.
We do not present these here, because they do not provide
any new insight.

Since we explicitly integrated over the distribution function (\ref{f-2-streams}) 
in deriving these results, we only give the transport equations 
(\ref{trans-eq-reg}) for the two-stream system in the Fokker-Planck form:
\be
\nonumber 
\big({\cal D}
- \nabla_p^i X^{ij}(t,{\bf v}) \nabla_p^j \big)
\left\{
\begin{array}{c}
n(t,{\bf r},{\bf p}) \cr
\bar n(t,{\bf r},{\bf p}) \cr
\end{array}
\right\}
= 0 \;, \qquad\qquad
\label{FP-eq-2-stream}
\big({\cal D} - \nabla_p^i X^{ij}_g(t,{\bf v}) \nabla_p^j \big)
n_g(t,{\bf r},{\bf p})
=  0\;,
\ee
where
\ba
\label{Xij-two-stream}
X^{ij}(t,{\bf v}) \equiv
\frac{g^4}{4}\, \frac{N_c^2-1}{N_c} \, n 
\int {d^3k \over (2\pi)^3} 
\frac{k^i k^j }{{\bf k}^4 (\omega_+^2 +\gamma_{\bf k}^2)^2}
\frac{\big(\gamma_{\bf k}^2 + ({\bf k} \cdot {\bf u})^2\big)^3} 
{\gamma_{\bf k} \big(\gamma_{\bf k}^2 + ({\bf k}{\bf v})^2\big)}
\, \sinh(2 \gamma_{\bf k}t) \;,
\ea
and $X^{ij}_g(t,{\bf v}) \equiv 2N_c^2 X^{ij}(t,{\bf p})/(N_c^2-1)$.
\end{widetext}

To get an idea how the two-stream system evolves according to the 
Fokker-Planck equations (\ref{FP-eq-2-stream}), we take into account
in the integral (\ref{Xij-two-stream}) only those wave vectors which
are parallel to the stream velocity ${\bf u}$, with the latter being chosen 
along the axis $x$. The only non-vanishing component of  
$X^{ij}(t,{\bf v})$ is then $X^{xx}(t,{\bf v})$. Neglecting the dependence 
of $X^{xx}(t,{\bf v})$ on ${\bf p}$ and assuming that the system 
is homogenous, the Fokker-Planck equation (\ref{FP-eq-2-stream})
for quarks becomes a one-dimensional diffusion equation
\be
\label{diff-eq}
\frac{\partial n(t,{\bf p})}{\partial t}
= D(t)\, \frac{\partial^2 n(t,{\bf p})}{\partial p_x^2} \;,
\ee
with the diffusion coefficient $D (t) \equiv X^{xx}(t)$
depending on time approximately as 
\be
\label{diff-const}
D (t) = d\, e^{2\gamma t} \;,
\ee
where $d$ and $\gamma$ are constants.

If the distribution function is initially of the form
\be
n(t=0,{\bf p}) = 2\pi \, \tilde{n} \, \delta (p_x - q) \,,
\ee
where $\tilde{n}$ is independent of $p_x$, the solution 
of the diffusion equation (\ref{diff-eq}) is found as
\be
\label{diff-eq-solution}
n(t,{\bf p}) = \tilde{n}
\sqrt{\frac{2\pi \gamma}{d(e^{2\gamma t}-1)}} \,  
\exp \bigg[-\frac{\gamma (p_x-q)^2}{2d(e^{2\gamma t}-1)} \bigg] \;.
\ee
The distribution function (\ref{diff-eq-solution}) is 
normalized in such a way that
$$
\int \frac{dp_x}{2\pi} n(t,{\bf p}) = \tilde{n} \;.
$$
According to the solution (\ref{diff-eq-solution}), the electric
field growing due to the electrostatic instability rapidly 
washes out the peak-like structures of the two-stream distribution 
function (\ref{f-2-streams}). It should be understood, however, 
that the solution (\ref{diff-eq-solution}) is valid only for time 
intervals which are sufficiently short that the distribution 
function used to compute the coefficient $X^{ij}(t,{\bf v})$ is 
not much different from the function (\ref{f-2-streams}). 
Nevertheless, the solution (\ref{diff-eq-solution}) shows 
how the equilibration process commences. 


\section{Summary and outlook}
\label{sec-discussion}


We have developed here the quasi-linear transport theory of a
weakly coupled quark-gluon plasma. Our main motivation was to study 
the equilibration of plasmas that are initially unstable. The field
fluctuation spectrum, which is found within the linear response 
approach, determines the evolution of the regular distribution functions. 
More specifically, the fluctuations of chromodynamic fields provide 
collision terms to the transport equations of the regular 
distribution functions. We have limited our considerations to 
longitudinal chromoelectric fields, as then the field correlation 
functions are known for both the isotropic and two-stream systems. 
The collision terms were found in either the Balescu-Lenard or 
Fokker-Planck form. In the case of an isotropic plasma we showed
how the system equilibrates when a small fraction of particles
has a different temperature than the bulk. 

The case of the two-stream system is more interesting. The 
Fokker-Planck equation could be approximately written as an 
equation of diffusion in momentum space. The diffusion coefficient, 
which is given by the chromoelectric fields for the two-stream 
instability, exponentially grows in time. We found the exact solution
of the diffusion equation, which showed that the peak-like 
structures in the parton momentum distribution dissolve rapidly.

In nonrelativistic plasmas it is often a well justified approximation
to keep only longitudinal electric fields and to neglect magnetic and 
transverse electric fields \cite{Ved61,Ved63}. In the case of
ultrarelativistic plasmas, this is no longer true. If initially the 
fields are purely longitudinal, the transverse fields are 
automatically generated, and they are dynamically important. 
Therefore, the ultrarelativistic plasma considered here, where the transverse 
fields are neglected, should be rather treated as a toy model which we 
have studied mostly for the sake of analytical tractability. With this 
simplified example we have been able to elucidate some general features 
of the problem. Physically better motivated situations will require
substantial numerical work, which is less conducive to general insights.

The considerations presented here clearly demonstrate the usefulness
of the quasi-linear transport theory for the study of equilibration
processes of quark-gluon plasmas. As mentioned in the Introduction,
numerical studies indicate that the unstable chromomagnetic 
plasma modes play an important role at the early stage of the 
quark-gluon plasma produced in relativistic heavy-ion collisions. 
Therefore, it would be of considerable interest to compute the
correlation functions of transverse fields in arbitrary anisotropic 
plasmas in order to derive the relevant transport equations.
As explained in \cite{Mrowczynski:2008ae}, there is no conceptual 
difficulty in such a computation, but one has to invert 
the matrix $\Sigma^{ij}(\omega,{\bf k}) \equiv
- {\bf k}^2 \delta^{ij} + k^ik^j 
+ \omega^2 \varepsilon^{ij}(\omega,{\bf k})$. This is easily 
done for isotropic plasmas but for anisotropic plasmas one obtains 
a rather complex expression which is very cumbersome for further 
analytic calculations \footnote{The problem greatly simplifies 
for longitudinal fields when, instead of the matrix 
$\Sigma^{ij}(\omega,{\bf k})$, one deals with the scalar 
function $k^ik^j\Sigma^{ij}(\omega,{\bf k}) 
= \omega^2 k^ik^j \varepsilon^{ij}(\omega,{\bf k})
= \omega^2 {\bf k}^2 \varepsilon_L(\omega,{\bf k})$.}. 
Except for some special cases, numerical methods seem to be 
unavoidable. Such computational studies are beyond the scope 
of the present work but progress in this direction will be 
hopefully reported soon.

\section*{Acknowledgments}

St.~M. is grateful to the Physics Department of Duke University, 
where this project was initiated, for warm hospitality during 
his visit. This work was supported in part by the 
U.~S.~Department of Energy under grant DE-F02-05ER41367.

\appendix*

\section{}


We discuss here the longitudinal chromodielectric permeability 
$\varepsilon_L(\omega,{\bf k})$ which is known to be
\be
\label{eL}
\varepsilon_L(\omega,{\bf k}) = 1+ \frac{g^2}{2{\bf k}^2}
\int {d^3p \over (2\pi)^3} 
\frac{{\bf k} \cdot \nabla_p f({\bf p})}
{\omega - {\bf k} \cdot {\bf v}+i0^+}
\;.
\ee
Applying the identity 
$$
\frac{1}{x \pm i0^+} = {\cal P}\frac{1}{x} \mp i\pi \delta(x)
$$
to Eq.~(\ref{eL}), one immediately finds $\Im \varepsilon_L(\omega,{\bf k})$ 
\be
\label{Im-eL}
\Im \varepsilon_L(\omega,{\bf k}) = - \frac{g^2}{4{\bf k}^2}
\int {d^3p \over (2\pi)^3} 
\, 2\pi \delta(\omega - {\bf k} \cdot {\bf v}) \, 
{\bf k} \cdot \nabla_p f({\bf p})
\;.
\ee
If the plasma is isotropic $\nabla_p f({\bf p})$ can be expressed as
\be
\label{nabla-f-iso}
\nabla_p f({\bf p}) = \frac{d f({\bf p})}{dE_p} \, {\bf v} \;.
\ee
And if the partons are additionally masslees, the integral in 
(\ref{eL}) factorizes into the angular integral and the integral 
over $p \equiv |{\bf p}|$. Then, one finds the real and imaginary parts 
of the longitudinal chromodielectric permeability 
$\varepsilon_L(\omega,{\bf k})$ as
\ba
\label{Re-Im-eL-massless}
\Re\varepsilon_L(\omega,{\bf k}) 
&=& 
1+ \frac{m_D^2}{{\bf k}^2}
\bigg[
1 - \frac{\omega}{2|{\bf k}|}
{\rm ln}\bigg|\frac{\omega + |{\bf k}|}{\omega - |{\bf k}|} \bigg| 
\bigg] \; ,
\nonumber \\
\Im\varepsilon_L(\omega,{\bf k}) 
&=& \frac{\pi}{2} \: 
\Theta ({\bf k}^2 -\omega^2) \:
\frac{m_D^2 \omega}{|{\bf k}|^3}  \; ,
\ea
where the Debye mass $m_D$ is 
\be
\label{m_D^2}
m_D^2 \equiv - \frac{g^2}{4\pi} \int_0^\infty dp \,p^2 
\frac{d f({\bf p})}{d p} \;.
\ee


\end{document}